\documentclass{appolb}
\usepackage{epsfig}

\begin{document}
\title{CP violation in the chargino/neutralino sector of the
MSSM\thanks{Presented by J. Kalinowski at the Epiphany Conference
on LHC Physics, 4 - 6 January 2008, Cracow, Poland.}%
}
\author{Jan Kalinowski and Krzysztof Rolbiecki\thanks{Supported by the
Polish Ministry of Science and Higher Education Grants
No.~1~P03B~108~30, No.~N N202 2161 33, and by the EU  Network MRTN-CT-2006-035505 ``Tools
and Precision Calculations for Physics Discoveries at Colliders".
 }
\address{Institute of Theoretical Physics, University of Warsaw\\
Ho\.{z}a 69, 00-681 Warsaw, Poland}} \maketitle
\begin{abstract}
The CP-violating effects in the neutralino sector of the MSSM can be
observed in event-counting type experiments, i.e.\ without the need
of exploiting variables sensitive to beam or neutralino
polarization. On the other hand, such CP-odd effects in the chargino
sector can be generated only at the loop level. We contrast the two
sectors and present results of full one-loop analysis of the  CP-odd
asymmetry in the non-diagonal chargino pair production in $e^+e^-$
collisions.
\end{abstract}

\section{Introduction}
Supersymmetry (SUSY) \cite{MSSM}, one of the most promising
extensions of the Standard Model (SM), introduces many new sources
of CP violation that may be needed to explain baryon asymmetry of
the universe. Experimental bounds on lepton, neutron and mercury
electric dipole moments (EDM)~\cite{susycp} restrict some of  the CP
phases to be vanishing or require internal
cancelations among various contributions~\cite{Ibrahim:2007fb}. In
the absence of any reliable theory of CP violation, however,
scenarios with some of the phases large and arranged consistent with
experimental EDM data have to be investigated. In such CP-violating
scenarios charginos and neutralinos (denoted generically by
$\tilde{\chi}$) might be light enough to be copiously produced at
hadron and lepton colliders, and thus provide information on CP
phases.

Many phenomena will be affected by non-vanishing CP phases:
sparticle masses, their decay rates and production cross sections,
SUSY contributions to SM processes etc. However, most unambiguous
way to detect the presence of CP-violating phases would be to study
CP-odd observables measurable at future accelerators -- the LHC and
the ILC. To build a CP-odd observable in a four-fermion process
(e.g.\ $e^+e^-\to\tilde{\chi}_i\tilde{\chi}_j$ or $\tilde{\chi}_i\to
f\bar f'\tilde{\chi}_j$) one typically uses spin information of one
of the particles involved. For example, a measurement of the fermion
polarization $s$ transverse to the production
plane~\cite{Kizukuri:1993vh} allows one to build a CP-odd observable
$s\cdot(p_e\times p_{\tilde{\chi}})$. This requires either
transverse beam polarization and/or spin-analyzer of produced
$\tilde{\chi}$'s via angular distributions of their decay
products~\cite{Gudi}. Another possibility is to look into triple
products involving momenta of the decay products of
charginos/neutralinos~\cite{Kittel:2004kd,Bartl:2008fu}.

However, CP-odd asymmetries can also be constructed from simple
event-counting type experiments if several processes are measured.
In the neutralino sector, due to their Majorana
nature, such a CP-odd asymmetry can be
constructed already at tree level~\cite{Choi:2001ww}-\cite{choi}, while in the
chargino sector it can be build at
one-loop~\cite{Osland:2007xw,Rolbiecki:2007se}.

In this contribution we contrast these two sectors and present a
CP-odd observable constructed beyond tree level from chargino
production cross section without polarized $e^+e^-$ beams or
measurement of chargino polarization. The CP asymmetry can be
induced by the complex higgsino mass parameter $\mu$ or complex
trilinear coupling in top squark sector $A_t$. Since the asymmetry
can reach a few percent, it can be detected in simple event-counting
experiments at future colliders.

\section{CP-violating effects in chargino and neutralino production at tree level}

In $e^+e^-$ collisions charginos are produced at tree-level via the
$s$-channel $\gamma,Z$ exchange and $t$-channel $\tilde{\nu}_e$
exchange, while the neutralino production receives contributions
from the $s$-channel $Z$ exchange and from both $t$- and $u$-channel
selectron exchanges.

\begin{figure}[tb]
\begin{center}
\epsfig{file=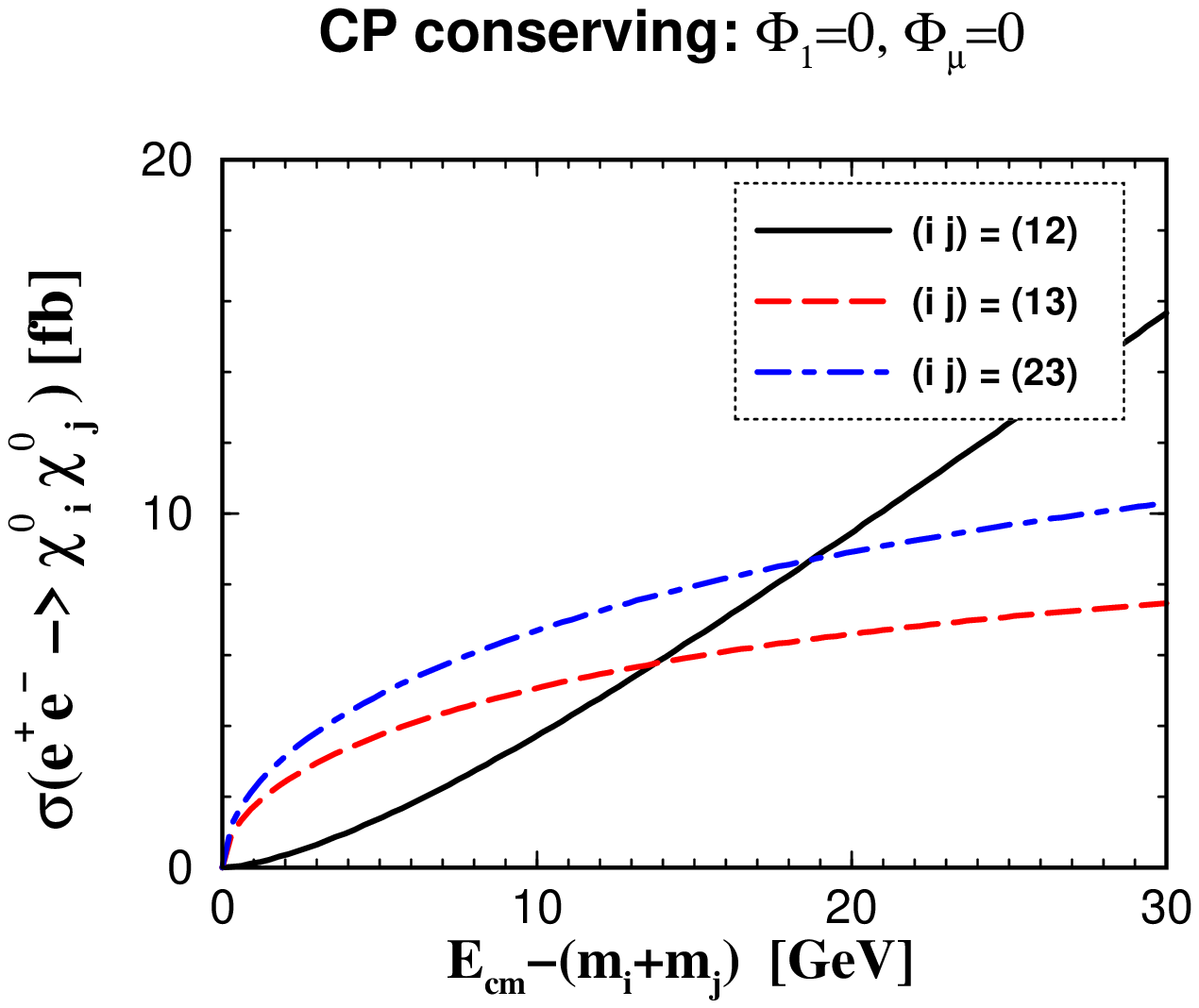,height=6cm,width=6cm}
\epsfig{file=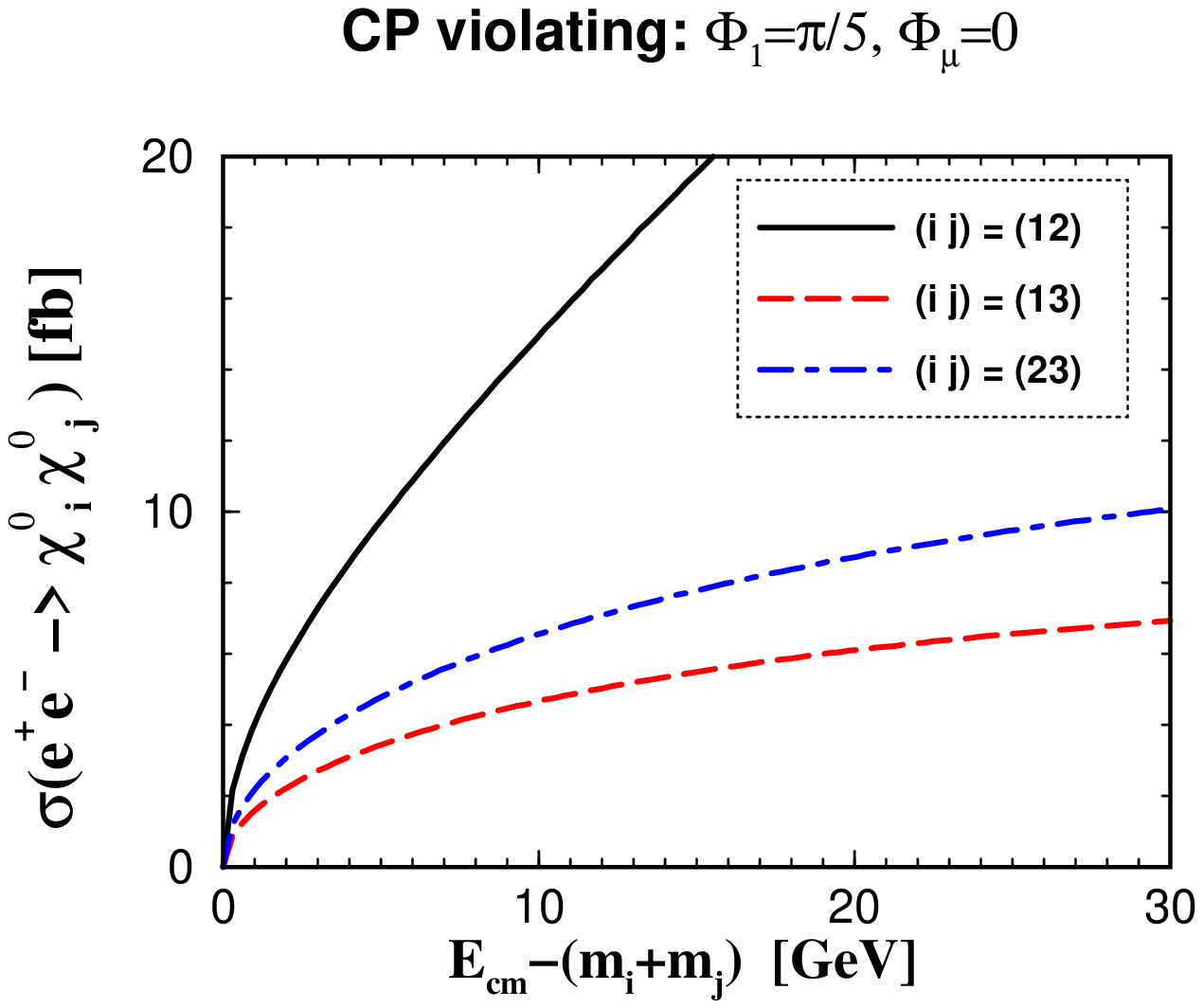,height=6cm,width=6cm}
\caption{\it The threshold behavior of the neutralino  production
  cross sections $\sigma^{\{ij\}}$ in the CP-conserving (left panel)
and the CP-violating (right panel) cases; $\{ij\}$ as indicated in
the figure (from~\cite{JKfest}).} \label{fig:th}
\end{center}
\end{figure}

Due to Poincar\'{e} invariance the unpolarized differential cross
section may depend only on masses $m_i, m_j$ and on two independent
scalar variables $s$ and $t$. As a result, the unpolarized
differential cross sections for equal-mass fermions $m_i=m_j$ in the
final state are always CP-even. However, if the fermion species in
the final state are different, beyond tree level the CP-odd
asymmetry can be built from the unpolarized cross sections for
non-diagonal chargino pair production $\sim \sigma(\tilde{\chi}^+_1
\tilde{\chi}^-_2) - \sigma(\tilde{\chi}^+_2
\tilde{\chi}^-_1)$~\cite{Osland:2007xw}, as we will discuss it in
the next section. Although the Majorana nature prevents to construct
a corresponding  CP-asymmetry for neutralinos, this same Majorana
nature opens up a possibility of investigating the CP violation
already at tree level by studying the energy behavior of the cross
sections for non-diagonal neutralino pair production near
thresholds~\cite{Choi:2001ww,JKfest,choi}.

In CP-invariant theories, the CP parity of a pair of Majorana
fermions $\tilde{\chi}^0_i\tilde{\chi}^0_j$ produced through a
vector or axial vector current with positive intrinsic CP parity is
given by
\begin{eqnarray}
1=\eta^i\eta^j (-1)^L
\label{cpparity}
\end{eqnarray}
in the non-relativistic limit, where $\eta^i$ is the CP parity of
$\tilde{\chi}^0_i$ and $L$ is the angular momentum \cite{11a}.
Therefore in $e^+e^-$ collisions neutralinos with the same CP
parities (for example for $i=j$) can be excited only in the P-wave.
The  S-wave excitation, with the characteristic steep rise $\sim
\beta$ (where $\beta$ is the neutralino c.m.\ velocity) of the cross
section near threshold, can occur only for non-diagonal pairs with
opposite CP parities of the produced neutralinos \cite{R2}.

If, however, CP is violated the angular momentum of the produced
neutralino pair is no longer  restricted by Eq.~(\ref{cpparity})
and all non-diagonal pairs are excited in the
S-wave~\cite{Choi:2001ww,JKfest}. This is illustrated in
Fig.\ref{fig:th}, where the threshold behavior of the neutralino
pairs $\{12\}$, $\{13\}$ and $\{23\}$ for the CP-conserving (left
panel) case is contrasted to the CP-violating case (right panel).
Even for relatively small CP phase $\Phi_1=\pi/5$, implying small
impact on CP-even quantities, the change in the energy dependence
near threshold can be quite dramatic. Thus, observing the  $\{ij\}$,
$\{ik\}$ and $\{jk\}$ pairs to be excited {\it all} in S-wave
would therefore signal CP violation.

\begin{figure}[htb]
\begin{center}
\epsfig{file=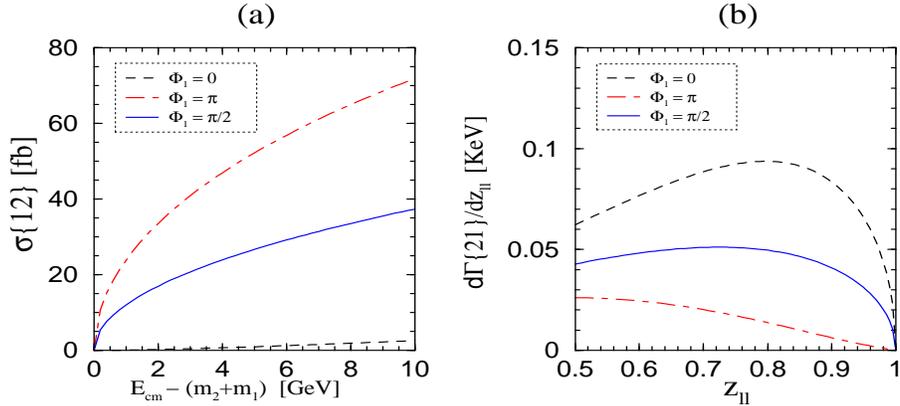,height=5.6cm,width=12cm} \caption{\it (a) The
         threshold behavior of the neutralino production
         cross sections $\sigma{\{12\}}$ and (b)
         the lepton invariant mass distributions of the decay
         $\tilde{\chi}^0_2\rightarrow\tilde{\chi}^0_1\, l^+l^-$ near the
         end point (from \cite{choi1}).}
\label{fig:proddecth}
\end{center}
\vskip -0.3cm
\end{figure}

Similarly the CP violation can be studied by investigating the
threshold behavior of the invariant mass distribution of two
fermions in the neutralino decay
$\tilde{\chi}^0_i\to\tilde{\chi}^0_j f^+f^-$ \cite{Choi:2001ww}. In
the CP-conserving case the decay amplitude satisfies the relation
\begin{eqnarray}
 1 =
-\eta^i\eta^j (-1)^L\,, \label{cpparity2}
\end{eqnarray}
in the non-relativistic limit of two neutralinos, where $L$ is the
orbital angular momentum of the final state of $\tilde{\chi}^0_j$
and fermion current. The relative minus sign  is a consequence of
two $u$-spinors for neutralino current in the decay amplitude as
compared to the $u$- and $v$-spinors in the production. The
immediate consequence of the selection rules (\ref{cpparity}) and
(\ref{cpparity2}) is that, in CP-invariant theories, if the
production of a pair of neutralinos with the same (opposite) CP
parity is excited slowly in P-wave (steeply in S-wave) in $e^+e^-$
collisions, then the neutralino to neutralino decay is excited
sharply in S-wave (slowly in P-wave). Therefore, the CP violation
can clearly be signalled by the simultaneous S-wave excitations of
the production of any non-diagonal $\{ij\}$ neutralino pair in
$e^+e^-$ annihilation near threshold and of the fermion invariant
mass distribution of the corresponding neutralino 3-body decay near
the end point. This is an interesting observation particularly if
only the light neutralinos $\tilde{\chi}^0_{1,2}$ among the four
neutralino states happen to be kinematically accessible in the
initial phase of $e^+e^-$ linear colliders. The combined analysis of
the threshold excitation of the production process,
$e^+e^-\rightarrow \tilde{\chi}^0_1 \tilde{\chi}^0_2$, and the
fermion invariant mass distribution of the decay,
$\tilde{\chi}^0_2\rightarrow\tilde{\chi}^0_1\, f\bar{f}$, near the
end point, as seen in Fig.~\ref{fig:proddecth}, can serve as one of
the most powerful probes of CP violation in the neutralino system
even in the initial phase of $e^+e^-$ linear colliders. The lepton
invariant mass distribution near the end point in the neutralino
decay alone can also be measured at the LHC providing information on
the relative CP parities of the two lightest neutralinos.

\section{CP-odd asymmetry in chargino production at one loop}
As shown in \cite{Choi:1998ei}, no CP-violation effects can be
observed at the tree-level for the non-diagonal $\tilde{\chi}_i^+
\tilde{\chi}_j^-$ chargino pairs without the polarization measurement of
final charginos. However the situation is different
if we go beyond tree-level approximation.

Radiative corrections to the chargino pair production include the
following generic one-loop Feynman diagrams: the virtual vertex
corrections, the self-energy corrections to the $\tilde{\nu}$, $Z$
and $\gamma$ propagators, and the box diagrams contributions. We
also have to include corrections on external chargino legs.

The one-loop CP asymmetry for the non-diagonal chargino
pair is defined as
\begin{eqnarray}
&& A_{12}=\frac{\sigma^{12}_{\rm loop}-
  \sigma^{21}_{\rm loop}}{\sigma^{12}_{\rm tree}+
  \sigma^{21}_{\rm tree}}\, ,
  \label{CPasym}
\end{eqnarray}
where $\sigma^{12}$, $\sigma^{21}$ denote cross sections for
production of $\tilde{\chi}_1^+ \tilde{\chi}_2^-$ and
$\tilde{\chi}_2^+ \tilde{\chi}_1^-$, respectively. The one-loop
corrected matrix element squared is given by
\begin{eqnarray}
|\mathcal{M}_{\mathrm{loop}}|^2 = |\mathcal{M}_{\mathrm{tree}}|^2 +
2 \mathrm{Re}(\mathcal{M}_{\mathrm{tree}}^*
\mathcal{M}_{\mathrm{loop}} )\, .
\end{eqnarray}
Since the asymmetry $A_{12}$ vanishes at tree level it has to be UV
finite. Since the structure of counter terms is the same as the tree
level graphs, using the same arguments as for the tree-level
amplitude it can be shown that renormalization procedure does not
give rise to the asymmetry. Nevertheless self-energy and vertex
corrections are UV-divergent, and proper treatment of divergences is
needed.

Loop diagrams with internal photon line also introduce infrared
singularities. They can be removed by adding emission of soft
photons from external charged particles. The sum of both
contributions is then IR finite, however it depends on the soft
photon cut. On the other hand soft photon emission part has the form
of tree-level amplitude multiplied by soft photon factor. Therefore,
the terms arising due to soft photon bremsstrahlung do not affect
the asymmetry $A_{12}$. Similar arguments apply for hard photon
emission from external fermions.

\begin{figure}[!t]
\begin{center}
\includegraphics[scale=0.365]{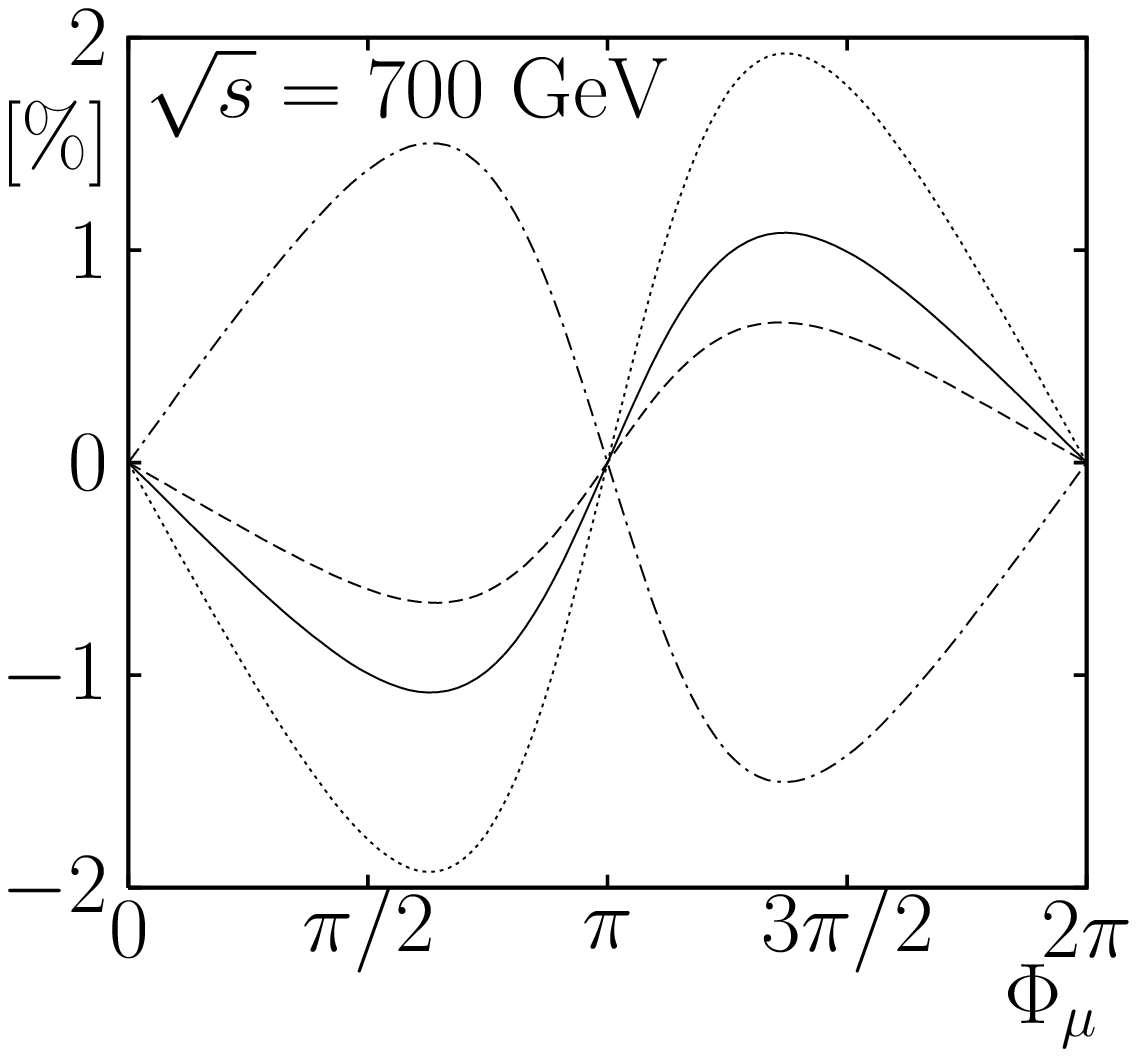}
\includegraphics[scale=0.35]{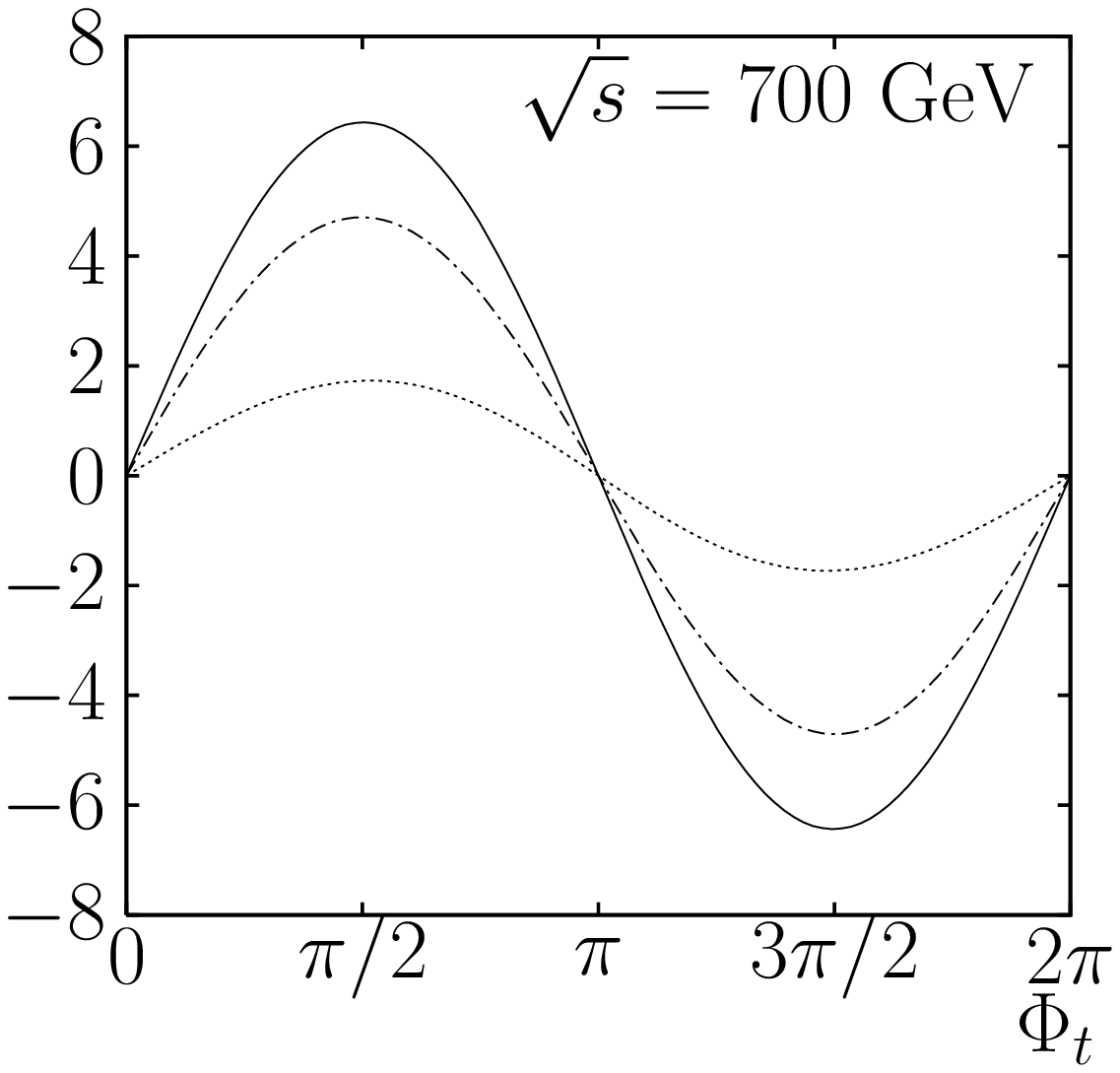}
\includegraphics[scale=0.365]{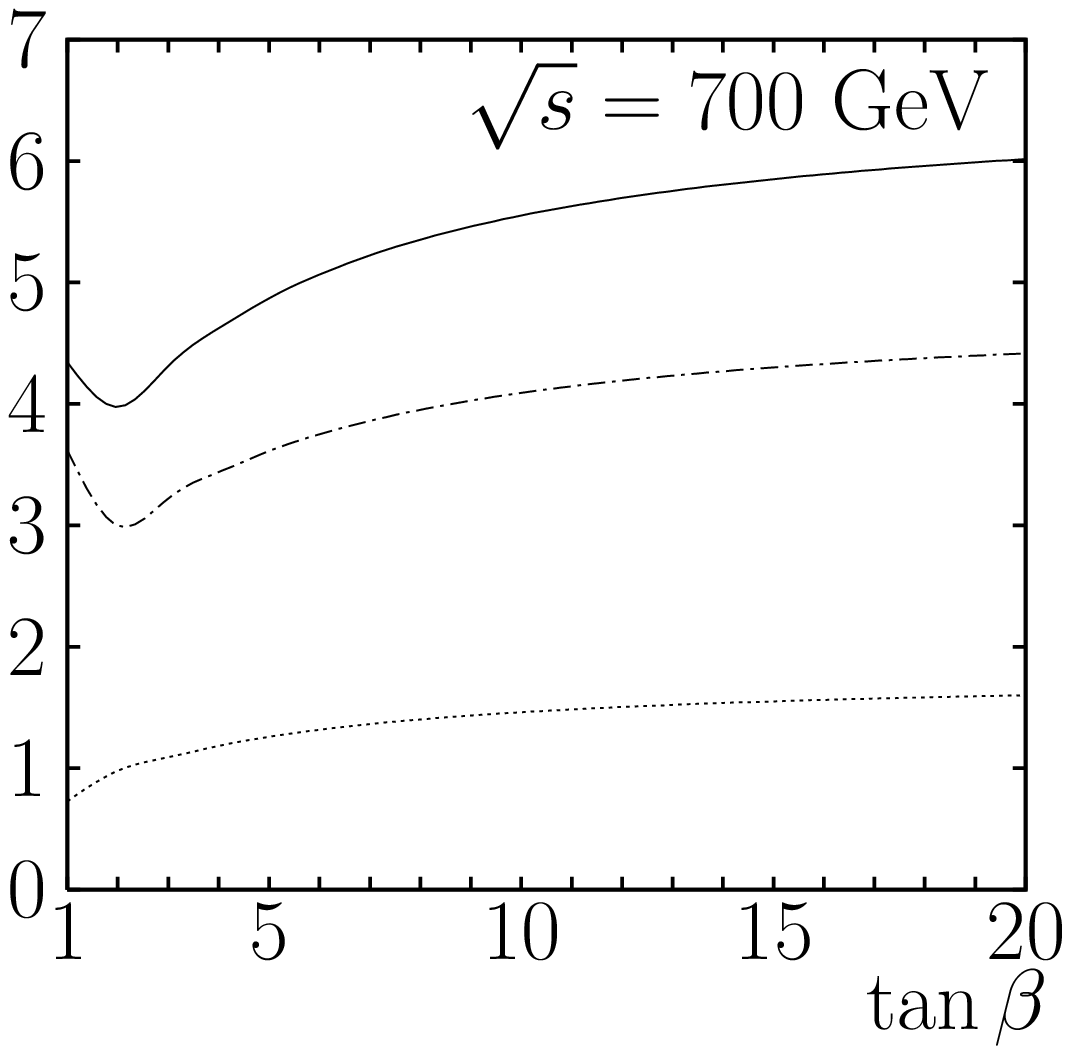}
\end{center}\vspace{-0.1cm}
\caption{\it Asymmetry $A_{12}$ as a function of the phase of $\mu$
parameter (left), the phase of $A_t$ (middle), and as a function of
$\tan\beta$ with $\Phi_t=\pi/3$ (right). Different lines denote full
asymmetry (full line) and contributions from box (dashed), vertex
(dotted) and self energy (dash-dotted) diagrams (from
\cite{Rolbiecki:2007se}). \label{fig2}}
\end{figure}

The CP asymmetry Eq.~(\ref{CPasym}) arises due to the interference
between complex couplings, which in our case appear in complex
mixing matrices of charginos or stops, and non-trivial imaginary
part from Feynman diagrams -- the absorptive part. Such
contributions appear when some of the intermediate state particles
in loop diagrams go on-shell. Therefore we take the following
scenario: gaugino/higgsino mass parameters $|M_1| = 100$ GeV, $M_2 =
200$ GeV, $|\mu| = 400$ GeV with $\tan\beta=10$. This gives the
following chargino masses: $ m_{\tilde{\chi}^-_1} = 186.7$  GeV,
$m_{\tilde{\chi}^-_2} = 421.8$ GeV. For the sfermion sector  we
assume universal slepton mass
$M_{\tilde{L}_{1,2,3}}=M_{\tilde{E}_{1,2,3}}=150$ GeV, while for
squarks $m_{\tilde{q}}\equiv
M_{\tilde{Q}_{1,2}}=M_{\tilde{U}_{1,2}}=M_{\tilde{D}_{1,2}}=450$ GeV
and $M_{\tilde{Q}}\equiv
M_{\tilde{Q}_{3}}=M_{\tilde{U}_{3}}=M_{\tilde{D}_{3}}=300$ GeV and
for the sfermion trilinear coupling:
$|A_{t}|=-A_{b}=-A_{\tau}=A=400\mbox{ GeV}$.

In our numerical analysis we consider the dependence of the
asymmetry (\ref{CPasym}) on the phase of the higgsino mass parameter
$\mu = |\mu| e^{i \Phi_\mu}$ and soft trilinear top squark coupling
$A_t = |A_t| e^{i \Phi_t}$. In Fig.~\ref{fig2} we show the CP
asymmetry as a function of the phase of $\mu$ and $A_t$, left and
middle panel, respectively.  Contributions due to box corrections,
vertex corrections and self energy corrections have been plotted in
addition to the full result. In this scenario the asymmetry can
reach $\sim 1\%$ for the $\mu$ parameter and $\sim 6\%$ for $A_t$,
respectively. We note that for the asymmetry due to the non-zero
phase of the higgsino mass parameter there are significant
cancelations among various contributions. In addition, we also show
in the right panel of Fig.~\ref{fig2} the dependence of the
asymmetry due to $A_t$ as a function of $\tan\beta$.

For the asymmetry generated by the $\mu$ parameter all possible
one-loop diagrams containing absorptive part contribute. The
situation is different for the phase of  the trilinear coupling
$A_t$ -- when chargino mixing matrices remain real. In this case
only vertex and self-energy diagrams containing stop lines
contribute to the asymmetry~\cite{Rolbiecki:2007se}.

\section{Summary}
It has been shown that the CP-odd observable can be constructed from
unpolarized chargino and neutralino production cross sections alone
without polarized beams nor the need of measuring
chargino/neutralino polarizations in the final state. While the
Majorana nature allows to probe the CP violation via the threshold
behavior of the production/decay rate, in the chargino sector the
CP-odd asymmetry arises at one loop and is generated by the
interference between complex couplings and absorptive parts of
one-loop integrals. The effect is significant for the phases of the
higgsino mass parameter $\mu$ and the trilinear coupling in the stop
sector $A_t$. At future colliders it may give information about CP
violation in chargino and stop sectors.

\end{document}